\magnification=1200
\hoffset=-.25in
\voffset=.0in
%
 
\vsize=8.0in 
\hsize=6in 
\tolerance 10000 
 
\baselineskip 12pt plus 1pt minus 1pt 
\ \ 
\bigskip 
\centerline{\bf Spectral Equivalence of Bosons and Fermions in}
\centerline{\bf One-Dimensional Harmonic Potentials}
\noindent          
\vskip .5in
\centerline{\it by}
\medskip
\centerline{\rm M. Crescimanno}
\smallskip
\centerline{\it Physics Department}
\centerline{\it Berea College}
\centerline{\it Berea, KY ~40404}
\medskip 
\centerline{\it and}
\medskip
\centerline{\rm A. S. Landsberg}
\smallskip
\centerline{\it W. M. Keck Science Center}
\centerline{\it The Claremont Colleges}
\centerline{\it Claremont, CA 91711}
\vskip .7in
\centerline{\it February, 2000}
\vskip 1.2in
{\bf ABSTRACT:} Recently, Schmidt and Schnack 
(Physica A260 479 (1998)), following 
earlier references, reiterate that the specific heat of $N$
non-interacting bosons in a one-dimensional harmonic 
well equals that of 
$N$ non-interacting fermions in the same potential. 
We show that this peculiar relationship between 
heat capacities results from a more dramatic equivalence 
between bose and fermi systems. 
Namely, we prove that the excitation spectrums 
of such bose and fermi systems are
{\it spectrally equivalent}. Two 
complementary proofs of this equivalence are provided, one 
based on an analysis of the dynamical 
symmetry group of the $N$-body system, 
the other using a combinatoric analysis.  

\vfill
\eject

\ \ 
\vskip .3in
\noindent {\bf I. Introduction:} With the advent of dilute atomic 
BEC$^{[1,2,3]}$
and, recently, nearly degenerate dilute atomic fermi gas$^{[4,5]}$, there is 
renewed interest in understanding aspects of quantum many-body theory in 
inhomogeneous (in particular, harmonically trapped) systems. 
Since trapped, cooled atoms have properties that are, 
in principle, controllable 
to a degree unavailable in other systems (e.g., clusters and nuclei), they  
present new opportunities to study quantum mechanics
and many-body theory. 

Although the ultra-cold dilute 
atomic gas systems are large compared to the coherence lengths,  
they are not homogeneous, due to the fact they are
generally trapped in a (nearly) harmonic potential. In 
many of these systems the interparticle forces are significant.  
In this note, however, we ignore the interactions between atoms, with the  
motivation being to understand better
the thermodynamic 
properties of $N$ trapped {\it non-interacting} bosons and fermions. 
Recent work$^{[6,7]}$ describes strange relations between
the equilibrium thermodynamics of these
two systems. 
It was shown, for example, that the heat 
capacity (as a function of temperature) 
of $N$ noninteracting bosons in a 
one-dimensional harmonic potential 
is the same as that of $N$ 
noninteracting fermions in an 
identical potential. The 
respective partition functions 
for these systems are likewise 
closely related (see Ref. [7]).

These "coincidences" provide hints that a deeper 
underlying connection exists between bose and fermi 
gases in a harmonic well. In particular, the heat capacity and partition
functions as functions of the inverse temperature $\beta$  
can be thought of as an
``imaginary time'' continuation of a fourier transform of the
spectrum. The fact that the heat capacities are the 
same for all temperatures 
suggests that there should be a state-for-state, level-for-level 
correspondence between these non-interacting 
many-body bosonic and fermionic systems. 
We show that this is indeed the case, 
and below describe two independent proofs of the
spectral equivalence of excitations in these systems. 
The first of these arguments relies on the dynamical 
symmetry group properties of these systems, 
while the other is based on combinatoric methods.
 
\bigskip
\noindent {\bf II. Dynamical symmetry group approach:}
 Consider $N$ non-interacting bosons or 
fermions in a harmonic well.  In either 
case, the energy differences between energy levels 
is an integer multiple of $\hbar\omega$.
Consequently, to analyze the spectral properties of each system 
it suffices to study the multiplicity per (total energy) 
level. 
Below we will show that the entire spectrum 
of both the bosonic and fermionic systems 
are isomorphic up to an overall energy shift. 

Classically, a system of $N$ non-interacting particles 
in a 1-d harmonic potential is identical to that of a single 
particle in an $N$-dimensional isotropic harmonic potential.
The system thus has an obvious spatial $O(N)$ symmetry we call
``angular momentum''. 
However, it is apparent with more 
introspection that the system possesses a much larger
dynamical symmetry group. Orbits in the $N$-dimensional 
isotropic harmonic potential do not precess. In analogy 
with the Kepler problem, we say that there is a conserved
Runge-Lenz vector (which may be thought of as the axis of the
orbit in configuration space), and we thus expect the 
symmetry group to be enlarged. 

Since we will be interested in the quantization of the
system, we describe the dynamical symmetry enlargement 
through the operators of the associated quantum theory.
To simplify notation, take $\hbar\omega=1$ throughout. 
Label the raising and lowering operators for the bosonic 
theory $a_i^{\dagger}, a_i$ with $i=1,\ldots, N$.
The canonical commutation relations (for the bosonic case) 
are  $[a_i, a_j^{\dagger}]=\delta_{ij}$.
The many-body hamiltonian operator of this noninteracting 
system is $H=\sum_i a_i^{\dagger}a_i + \epsilon$, where
$\epsilon$ is an overall constant. 

We call the space of the eigenvalues of the
$a_i^{\dagger}a_i$ the state space. Equivalently, the 
state space is the integer lattice in the 
$(+,\ldots,+)$ quadrant of $N$-dimensional 
Euclidean space. State space 
is not the Fock space, but is 
a useful auxiliary space from which we will 
construct the Fock space, and so we discuss it's properties. 
Let $e_i$ be orthonormal unit basis vectors in this Euclidean space
associated with the eigenvalues of $a_i^{\dagger}a_i$.
We name several distinguished vectors in this space, namely
the {\it level} vector
$k=\sum_i e_i$ and the {\it root} vectors
$l_i = e_i-e_{i+1}$ for $i=1, \ldots, N-1$.
We also define a spanning set of {\it weight} vectors $r_i$ via
$(r_i, l_j) = \delta_{i,j}$ together with $(r_i,k)=0$. 

Note that the operators associated with $l_i$, namely 
$a_i^{\dagger}a_i-a_{i+1}^{\dagger}a_{i+1}$, are independent, and commute 
with each other (being all diagonal) and with the hamiltonian. 
The hamiltonian corresponds to the level
vector.
Furthermore, to each pair of particles $l \neq j$, there is an associated
$su(2)$ subalgebra generated by $\{a_l^{\dagger}a_l-a_j^{\dagger}a_j,
a_l^{\dagger}a_j+a_j^{\dagger}a_l, i(a_l^{\dagger}a_j-a_j^{\dagger}a_l)\}$.
The application of the second or third operators in the above $su(2)$ subalgebra
``shift'' the first operator's eigenvalue by a combination of root vectors.
Finally, note that the matrix of inner products
$M_{ij} = (l_i,l_j)$ of the root vectors is exactly the 
cartan matrix of $su(N)$. Thus, we have identified the dynamical symmetry
group of this system generated by the (trace-free part of the) 
products of $a_i^{\dagger}a_j$
to be $su(N)$. 

We now construct the Fock space for both fermions and bosons 
from the state space 
by realizing the respective anti-symmetrizations and symmetrizations
of the multi-particle Fock states as 
linear combinations of states in the state space that lie 
on the same Weyl group orbit. 
We make this correspondence precise with the following observations. 
Each state in the state space 
can be thought of as a particular product of single particle states, 
its coordinates (the components of the $l_i$ are integers) are simply 
the harmonic oscillator level of each 
particle. Constructing the multi-particle state associated with 
that product of single particle states consists of combining 
all the states from single-particle label permutations. The permutation 
group $S_N$ is generated by primitive transpositions
$(\ldots, n_i, n_{i+1}, \ldots) \rightarrow 
(\ldots, n_{i+1}, n_i, \ldots)$. Each of these primitive transpositions
acts as a Weyl reflection (acting on all the roots) about the 
hyperplane perpendicular 
to the root $l_i$. 

Thus, the Weyl group, ${\cal W}$ of the symmetry 
algebra $su(N)$ is exactly the group of permutations 
of the single particle states that make up the many-body state. 
Each element of the Weyl group preserves the level $k$. 
We specify a many body state through an assignment of 
a highest weight vector $r$ and level $s$ (a natural number) for which 
$r+{{sk}\over{N}}$ is a vector in the 
$(+,\ldots,+)$ quadrant (boundaries included). 
Explicitly, in terms of the vectors in the state space, 
the bosonic many-body Fock space has the basis $\Psi^{\mit boson}_{r,s}$   
$$ \Psi^{\mit boson}_{r,s}={{1}\over{\sqrt{N!}}}\sum_
{\sigma \in {\cal W}} |{{s}\over{N}}k+\sigma\cdot r> 
$$
whereas the basis of the fermionic many-body Fock space is 
$$ \Psi^{\mit fermion}_{r,s}={{1}\over{\sqrt{N!}}}\sum_{\sigma \in {\cal W}} 
(-1)^{sgn(\sigma)}|{{s}\over{N}}k+\sigma\cdot r> 
$$
where the $sgn(\sigma)$ is 1 if $\sigma$ is an even permutation and 
-1 if it is an odd permutation. Note that according to this 
definition, only $r$ vectors from the interior of the Weyl chamber are 
associated with a fermionic many-body state. 

Succinctly stated, the multi-particle permutation symmetry of
quantum mechanics maps the single particle states of state space into
the highest weight space of the 
symmetry algebra $su(N)$. For bosons, the map covers the 
entire Weyl chamber (including the lattice points in the 
bounding hyperplanes) at each level. For fermions, 
the map covers only the interior
lattice points of the Weyl chamber. Additionally, due to the 
constraint that $r+{{sk}\over{N}}$ is in the 
$(+,\ldots,+)$ quadrant, at each 
level there are of course only a finite number of highest weight 
candidates. 

The vector $\rho = {{1}\over{2}}\sum_{\alpha>0} \alpha$
(half the sum of positive roots) 
translates the vacuum of the bosonic Fock space to that of the 
fermionic Fock space at each level. Note also that $\rho$ is 
thus orthogonal to the level vector $k$. 
It can be combined with the 
level vector to 
constitute a one-to-one map between the spectrum of the $N$ boson and 
$N$ fermion systems. Note that translation by the 
vector vector $\Gamma = (0,1,2, \ldots, N-1)$ is precisely that map, and that
$\Gamma = \rho+{{N-1}\over{2}} k$. Note further that $\Gamma$ has level 
$\Gamma\cdot k = N(N-1)/2$, which is precisely the ground state energy 
shift between the bosonic and fermionic system. Geometrically, $\Gamma$ is the 
smallest lattice vector that translates the lattice points in the bounding 
hyperplanes entirely into (the subset of) 
the interior of the Weyl chambers at each level. 

\bigskip
\noindent {\bf III Combinatoric approach:} 
The spectral equivalence of one-dimensional, 
noninteracting harmonically-trapped bosonic and fermionic gases 
can also be understood through a straightforward combinatoric
argument.

In a system of $N$ noninteracting particles (bosons or fermions)
 in a harmonic well let the energy level
of the $i$th particle be specified by the integer $e_{i}$, 
with $E=\sum_{i=1}^{N} e_i$ 
the total energy of the system. (Note: In writing the energy $e_{i}$ 
as an 
integer, we are, as before, setting $\hbar\omega=1$, and for 
notational convenience are ignoring the constant $1/2$ 
associated with the single-particle ground state energy.)   
Clearly there are many different microconfigurations
possessing the same total energy $E$; we let
$G_{N}(E)$ denote the multiplicity of states with fixed energy $E$. We will show that
the multiplicity functions for bosons and fermions are equivalent.  
More precisely, we show that
$G_{N}^{boson}(E)=G_{N}^{fermion}(E+{{N(N-1)}/{2}})$,
indicating that the multiplicities for the bose and fermi 
cases are identical provided each is measured 
relative to its respective ground state energy (i.e.,
$0$ for bosons and ${{N(N-1)}/{2}}$ for fermions).  
 This equivalence  is sufficient
to prove the spectral equivalence of the excitation spectrum.

We begin with the bose case. We imagine ordering the 
$N$ particles from lowest energy to highest
$(e_1, e_2, \ldots , e_N)$.
The energy of the lowest-energy particle ($e_1$) can range 
from zero up to a maximum value of $[E/N]$, where
the brackets $[ \;\; ]$ denote the integer part of 
the expression enclosed. (It is readily seen that 
if the energy of the {\it lowest}-energy particle 
were to exceed this maximum value, then the sum of 
the energies of the $N$ individual particles would 
exceed the total specified energy $E$ of the system.) 

For a fixed $e_1$, the remaining energy 
$E-e_1$ must be
divided up among $N-1$ particles. So the possible 
values of $e_2$, which represents the lowest 
energy among the remaining $(N-1)$ particles, 
can range  from $e_1$ to 
$\left[ {{E-e_1}\over{N-1}} \right] $.
(As before, it is clear that if $e_2$ 
went outside this range, then the sum of the 
energies of the $N-1$ particles would exceed the 
prescribed value $E-e_1$.)

Proceeding in this fashion, we see that 
$$ G_{N}^{boson}(E) = \sum_{e_1=0}^{[E/N]} 
\sum_{e_2=e_1}^{\left[ {{E-e_1}
\over{N-1}} \right]} \ldots \sum_{e_{N-1}
=\epsilon_{N-2}}^{\left[ {{E-e_1-e_2 - 
\ldots - e_{N-2}}\over{2}} \right] } 1 .$$

A similar argument is used to construct the multiplicity 
function for the fermionic case. The
fundamental distinction stems from the additional 
constraint that two fermions cannot occupy the same energy
orbital, which in turn modifies the lower and upper 
bounds in the above summations, as we now describe.
Consider first the lower bounds. From the exclusion 
principle, it immediately follows 
that the lower (fermionic) bounds must  take the form 
$e_i=e_{i-1} +1$. The upper limits are found by noting that
for a system of $N$ fermions with total energy $E$, the energy of the
lowest-energy fermion cannot exceed
$\left[{{E-{{N(N-1)}\over{2}}}\over{N}}\right]$, as a 
straightforward calculation reveals.  Consequently, we find
$$G_{N}^{fermion}(E)=
\sum_{e_1=0}^{\left[{{E-{{N(N-1)}\over{2}}}\over{N}}\right]}
\sum_{e_2=e_1+1}^{\left[{{E-e_1 
- {{(N-1)(N-2)}\over{2}}}\over{N-1}}\right]}
\ldots
\sum_{e_{N-1}=e_{N-2}+1}^{\left[{{E-e_1 
-e_2 - \ldots -e_{N-2}-{{(2)(1)}\over{2}}}
\over{2}}\right]} 1 .$$

Expressed in this manner, the equivalence of $G_{N}^{boson}(E)$ 
and $G_{N}^{fermion}(E+{{N(N-1)}/{2}})$ is now revealed  through 
the following key coordinate transformation: In the 
fermionic summations above, introduce 
new coordinates $\hat{e}_{i} = e_i -i+1$. 
We claim that this will transform the fermionic sum into the corresponding
bose sum. 
(Note: in the context of the preceding analysis, this
coordinate change serves to relate the interior lattice 
points of the Weyl chamber (fermionic case) to
the entire Weyl chamber (bose case), that is, it is simply the 
translation by the vector $\Gamma$)
To see that this transformation achieves the 
desired result, first observe that under this transformation, 
the lower bounds in the fermionic summations 
$(e_{i+1}=e_{i}+1)$ become 
$(\hat{e}_{i+1} = \hat{e}_{i})$, 
just as in the bose case. Meanwhile, it is not 
difficult to verify that the upper limits in the fermionic summations 
$$e_{i+1}=\left[{{E-e_1 -e_2 - \ldots 
-e_{i}-{{(N-i)(N-i-1)}\over{2}}}\over{N-i}}\right]$$ 
now take the form 
$$\hat{e}_{i+1}=\left[ {{E-{{N(N-1)}\over{2}}-\hat{e}_1-\hat{e}_2 
- \ldots - \hat{e}_{i}}\over{N-i}} \right],$$ 
which, again is the same as for the bosonic case 
(once we shift by the fermionic 
ground state energy $E \rightarrow E+N(N-1)/2$).

This equivalence between the bosonic 
and fermionic multiplicity functions proves that 
the excitation spectrum
of one-dimensional harmonically trapped $N$ non-interacting bosons 
is identical to that of $N$ non-interacting fermions.

\bigskip
\noindent {\bf IV. Remarks and Conclusion:} Although the excitation spectra
of the fermi and bose systems are identical, 
these systems are not related by an obvious 
supersymmetry. There may, however, exist a 
connection associated with the
fermionic representation of affine lie 
algebra characters as described in Refs.[9,10,11]. 
Lastly, we observe that the recent work of 
Schmidt and Schnack$^{6,7}$  indicates that 
the specific heats of  similar bose and 
fermi systems in higher spatial dimensions 
(specifically, odd dimensions) might also 
be equivalent, just as for the one-dimensional 
case considered here. However,  preliminary work suggests that
spectral equivalence does not persist in higher (odd) dimensions. 

\bigskip
\noindent{\bf V: Acknowledgments} This research was supported 
in part by Research Corporation 
Cottrell Science Award \#CC3943 and in part by the National Science 
Foundation under grants PHY 94-07194  and EPS-9874764.

\vfill 
\eject
\ \ 
\vskip .3in 
\centerline{\bf Bibliography}
\bigskip
\item{1.} M. H. Anderson, J. R. Ensher, M. R. Mathews, C. E. Weiman and 
E. A. Cornell, {\it Science} {\bf 269}, 198 (1995).
\medskip
\item{2.} K. B. Davis {\it et. al.}, {\it Phys. Rev. Lett.} 
{\bf 75}, 3969 (1995).
\medskip
\item{3.} C. C. Bradley, C. A. Sackett and R. G. Hulet, (to be published).
\medskip
\item{4.} B. DeMarco and D. S. Jin,  {\it Phys. Rev. A} {\bf 58}, R4267 (1998).
\medskip
\item{5.} B. DeMarco, Bohm, Burke, Holland, D. S. Jin, {\it Phys. Rev. Lett.},
{\bf 82}:(21) 4208, (1999), cond-mat/9812350.
\medskip
\item{6.} H.-J. Schmidt and J. Schnack, ``Thermodynamic fermion-boson symmetry
in Harmonic Oscillator Potentials,'' cond-mat/9810036
\medskip
\item{7.} H.-J. Schmidt and J. Schnack, Physica 
{\bf A260}  479, (1998) cond-mat/9803151
\medskip
\item{8.} J. E. Humphreys, ``Introduction to Lie 
Algebras and Representation Theory,''
Springer-Verlag, New York, 1972
\medskip
\item{9.} R. Kedem, T. R. Klassen, B. M. McCoy, and E. Melzer, 
{\it Phys. Lett.} {\bf B307} (1993) 68-76, hep-th/9301046
\medskip
\item{10.} E. Melzer, {\it Int. J. Mod. Phys.} {\bf A9} (1994), 1115-1136, 
hep-th/9305114
\medskip
\item{11.} E. Bauer and D. Gepner, {\it Phys. Lett.} {\bf B372} (1996) 231-235, 
hep-th/9502118

\par 
\vfill
\end